\documentclass[aps,prb,twocolumn,
				amsmath,amssymb,amsfonts,floatfix,
longbibliography,superscriptaddress,nobibnotes]{revtex4-2}

\usepackage{graphicx} 
\usepackage{bm} 
\usepackage{ulem}
\usepackage{physics}
\usepackage[hypertexnames=false,colorlinks=true,citecolor=blue,
			linkcolor=blue,urlcolor=blue]{hyperref}

\usepackage{xfrac}

\usepackage{xcolor}
\definecolor{mygreen}{rgb}{0.0, 0.6, 0.0}

\def\*#1{\mathbf{#1}} 

\def\cdag{c^{\dagger}}

\graphicspath{{./figs/}}

\begin{document}

\title{Kagome edge states under lattice termination,\\ spin--orbit coupling, and magnetic order}

\author{Sajid Sekh}
\email[e-mail: ]{sajid.sekh@ifj.edu.pl}
\affiliation{\mbox{Institute of Nuclear Physics, Polish Academy of Sciences, W. E. Radzikowskiego 152, PL-31342 Krak\'{o}w, Poland}}

\author{Annica M. Black-Schaffer}
\affiliation{\mbox{Department of Physics and Astronomy, Uppsala University, Box 516, S-751 20 Uppsala, Sweden}}

\author{Andrzej Ptok}
\email[e-mail: ]{aptok@mmj.pl}
\affiliation{\mbox{Institute of Nuclear Physics, Polish Academy of Sciences, W. E. Radzikowskiego 152, PL-31342 Krak\'{o}w, Poland}}

\date{\today}

\begin{abstract}
We study the edge state properties of a two-dimensional kagome lattice using a tight-binding approach, focusing on the role of lattice termination, spin--orbit coupling, and magnetic order. In the pristine limit, we show that the existence of localized edge states is highly sensitive to boundary geometry, with certain terminations completely suppressing edge modes. Kane--Mele spin--orbit coupling opens a bulk gap and stabilizes topologically protected helical edge states, yielding a robust $\mathbb{Z}_2$ insulating phase that is insensitive to termination details. In contrast, the combined effect of a Zeeman field and Rashba spin--orbit coupling drives the system into Chern insulating phases, with Chern numbers consistent with the number of chiral edge modes. We further demonstrate that non-coplanar magnetic textures generate multiple Chern phases through finite scalar spin chirality, with Kane--Mele coupling strongly tuning the topological gaps. Our results provide important insights into the tunability of edge states in the kagome lattice, which can be key to designing materials with novel electronic properties and topological phases.
\end{abstract}

\maketitle

\section{Introduction}
\label{sec:intro}

Edge states refer to low-dimensional electronic states that are pinned to the boundaries of a material. They play a central role in topological condensed matter.
In fact, the bulk-boundary correspondence~\cite{hatsugai.93}, a foundational principle in topological physics, dictates that nontrivial bulk invariants manifest as gapless modes confined to system boundaries. Edge states can thus serve both as a signature of nontrivial bulk topology and as functional channels for robust, dissipation-resistant transport.

At the same time, two-dimensional (2D) kagome lattice has emerged as a particularly rich platform for exploring novel phenomena. The structure of the kagome lattice is formed by corner-sharing triangles arranged in a hexagonal fashion, which provides geometrical frustration, preventing spins from aligning magnetically. This leads to exotic kagome spin ice states~\cite{matan.grohol.06,moessner.sondhi.01}, emergent gauge fields~\cite{hermele.fisher.04}, and fractionalized excitations~\cite{balents.fisher.02,han.helton.12}. Apart from spins, the kagome lattice is also exciting from an electronic perspective as the band structure contains Dirac cones, van Hove singularities (VHS), and flat bands (FB) at various fillings. The existence of a large density of states around VHS and FB can produce strong correlation effects. These features underpin a variety of exotic phases such as charge~\cite{teng.chen.22,tan.liu.21,ptok.kobialka.22}, spin~\cite{park.ortiz.25}, and pair density waves~\cite{chen.yang.21}, nematicity~\cite{yang.ye.24,asaba.onishi.24,nie.sun.22,xu.ni.22}, skyrmionic texture~\cite{hirschberger.nakajima.19,hirschberger.szigeti.24}, loop currents~\cite{mielke.das.22}, and unconventional superconductivity~\cite{ortiz.teicher.20,ortiz.sarte.21,guguchia.mielke.23} in kagome materials.

Edge states are an exciting avenue for unlocking the potential of the kagome lattice. In a pristine lattice system, the properties of edge states often strongly depend on the details of the boundary, making lattice termination a crucial factor. This dependence is already well-known in the honeycomb lattice of graphene~\cite{nakada.fujita.96,fujita.wakabayashi.96,son.cohen.06,fuente.carrascal.23}, where the nature of the edge dramatically influences the electronic spectrum: zigzag termination supports localized, nearly flat edge states at half filling, which give rise to enhanced local density of states and magnetism, whereas armchair termination lacks such modes. The sensitivity to termination has also been shown to be crucial for building quantum devices~\cite{guo.lin.09,li.zhou.13,zakharova.mastalygina.21}, and we can expect an even richer scenario in kagome lattices due to their corner-sharing triangular geometry. In fact, experimentally, lattice termination has already emerged as a viable tool to design and manipulate edge states in kagome materials. For instance, in the magnetic kagome material Co$_3$Sn$_2$S$_2$~\cite{mazzola.enzer.23}, different termination planes produce contrasting surface states near the Fermi level. Similarly, a ``terrace''-like geometry in antiferromagnetic FeGe~\cite{yin.jiang.22,liu.yang.24}, where kagome layers form a staircase pattern with distinct edge boundaries at each step, has revealed spin-polarized edge states around the Fermi level, further enhanced by charge order. Although some effort~\cite{sun.chen.22} has been made to understand the pristine case, it is not exhaustive, and in particular, does not answer how the edge states intertwine with spin-orbit coupling (SOC) and magnetism.

In fact, kagome materials are known to possess finite SOC, which can substantially enhance Berry curvature to produce nontrivial  topological edge states. Topological edge states are fundamentally different from termination-dependent trivial edge states. For example, trivial edge states can be removed or significantly modified by boundary conditions, such as surface potentials, dangling bonds, or lattice termination. In contrast, topological edge states originate from a nonzero topological invariant defined in the bulk, and as a result, these states are robust against any perturbations that do not close the bulk gap or break the protecting symmetry. Such immunity is highly promising in device applications, where disorder and imperfections are ubiquitous, and underpins the strong interest in topological edge states for fault-tolerant quantum computing architectures~\cite{kitaev.02,nayak.simon.08,sarma.freedman.15,aasen.aghaee.25}. It may even be that SOC effects turn trivial edge states into topologically protected states, protected by a nonzero bullk invariant and thus insensitive to boundary details. The role of SOC in generating topological phases is elucidated in the Kane--Mele model~\cite{kane.mele.05}, where spin-dependent next-nearest-neighbor hopping and a staggered potential give rise to a $\mathbb{Z}_2$ topological insulating phase in the honeycomb lattice.
Although formulated as a toy model, the underlying mechanism extends to kagome lattices. For example, first-principle analysis of the nonmagnetic kagome material $\mathrm{KV_3Sb_5}$~\cite{ortiz.sarte.21} has revealed a $\mathbb{Z}_2$ topological phase in the normal state, where the $\mathbb{Z}_2$ nature is confirmed by the wavefunction parity at time-reversal invariant momenta. This $\mathbb{Z}_2$ phase remains intact as long as the time-reversal symmetry (TRS) is present.

An external magnetic field is well-known to break TRS. More interestingly, TRS can also be broken by intrinsic magnetism or chiral charge order. The breaking of TRS can induce a topological Chern phase, characterized by a nonzero Chern number, which underlies the quantum anomalous Hall (QAH) effect. In the QAH state, a transverse Hall current arises purely from the Berry curvature of the electronic bands, without the need for an applied magnetic field. In magnetic kagome systems, the QAH phase has so far been found to be driven by ferromagnetism, with experimental signatures, including anomalous transport and neutron diffraction, having confirmed ferromagnetic order in Fe- and Co-based kagome compounds~\cite{liu.sun.18,tanaka.fujisawa.20,ye.kang.18}, as well as in the kagome-based 166 family~\cite{xu.yin.23,zhou.lee.24}. In addition, scanning microscopy results of an Fe-based kagome compound~\cite{yin.jiang.22} have revealed pronounced edge states within the bulk gap. The ferromagnetism can be effectively modeled as an out-of-plane Zeeman field. Such a Zeeman field produces spin--split bands, analogous to the scenario in 1D Majorana nanowires~\cite{lutchyn.sau.10}, which have been proposed~\cite{nayak.simon.08} as a platform for fault--tolerant~\cite{shor.96} quantum computation.

The QAH can also be realized with more complex magnetic orders beyond ferromagnetism. Unlike simple ferromagnets or antiferromagnets, spins on the kagome lattice often form noncollinear arrangements, with the planar $120^{\circ}$ ``$\mathbf{q}=0$'' configuration~\cite{chubukov.92,harris.kallin.92,gotze.farnell.11,watanabe.araki.22} being the simplest example. While this coplanar order does not break TRS, a non-coplanar canting of the spins has been shown to generate a finite scalar spin chirality, which can induce a QAH~\cite{taillefumier.canals.06} phase. Such canting can arise naturally from Dzyaloshinskii--Moriya interactions~\cite{elhajal.canals.02} or at interfaces in thin films, producing an “umbrella”-like spin structure. This non-coplanar order opens a topological bulk gap, but although the QAH phase has been predicted~\cite{taillefumier.canals.06}, the edge state dispersion and interplay with spin--orbit coupling remain largely unexplored for such magnetic configurations. Understanding these will allow us to have more precise and tunable control of non-trivial edge states in a realistic setup.

Motivated by the extensive work on kagome materials, our work aims to broaden the understanding of edge states in kagome systems. For the pristine case, we identify four distinct terminations arising from a line cut. We further demonstrate that edge modes vanish for one of these terminations similar to armchair honeycomb system, which has not been previously reported. While the effects of SOC and magnetic order have largely been explored in bulk contexts or through material-specific tight-binding models~\cite{bolens.nagaosa.19,sun.chen.22,fang.ye.22,titvinidze.legendre.22,watanabe.araki.22}, we systematically examine their influence on edge states across all terminations. Particular attention is devoted to the interplay between SOC and magnetic order. Notably, we demonstrate how SOC can be used to tune the topological phase induced by non-coplanar magnetism.

We organize the work as follows. We begin by discussing the bulk features of the kagome lattice in Sec.~\ref{sec:bulk}. We present the slab band structure and LDOS to show the effects of lattice termination in Sec.~\ref{sec:slab_termination}. We present results including Kane--Mele SOC in Sec.~\ref{sec:soc}, followed by a discussion of effective Zeeman field and Rashba SOC in Sec.~\ref{sec:qah}. We consider the non-coplanar magnetic texture in Sec~\ref{sec:ncl}. Finally, we conclude our work in Sec.~\ref{sec:summary}.

\begin{figure*}[t]
	\includegraphics[width=\linewidth]{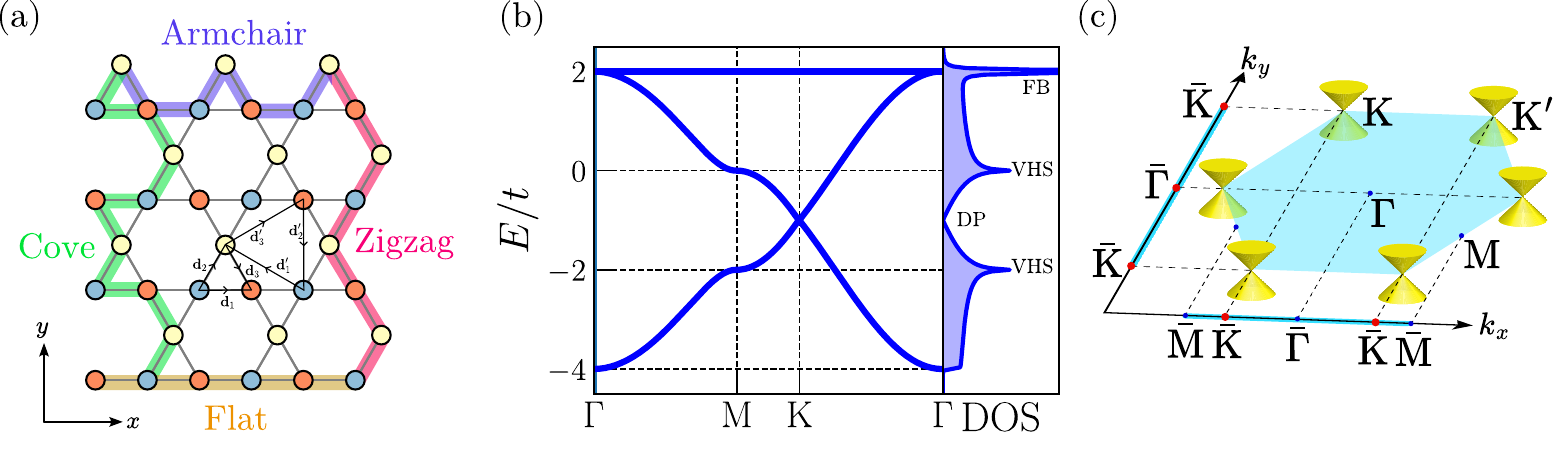}
	\caption{(a) Kagome lattice, schematically highlighting the four edge types arising from termination by a line cut. Solid arrows indicate nearest $\mathbf{d}_{1,2,3}$ and next-nearest $\mathbf{d}^{\prime}_{1,2,3}$ neighbor vectors, while site color-coding refers to the three sublattice atoms of the kagome lattice. (b) Band structure (left) and density of states (DOS) (right) in the kagome lattice, with features of Dirac point (DP), van Hove singularity (VHS), and flat band (FB) appearing at different fillings. (c) Full (blue hexagon) and projected slab (blue line) Brillouin zones of the kagome lattice. In contrast to the full Brillouin zone, the high symmetry points in the slab are written with bar. Yellow cones denote the Dirac points.}
	\label{fig:schematic}
\end{figure*}

\section{Key properties of the kagome band structure}
\label{sec:bulk}
We start with a single-orbital tight-binding description of the kagome lattice [see Fig.~\ref{fig:schematic}(a)]
\begin{eqnarray}
	H_{\text{KIN}} = -t \sum_{\langle ij \rangle \sigma} c^{\dagger}_{i\sigma} c_{j\sigma} - \mu \sum_{i\sigma} c^{\dagger}_{i\sigma} c_{i\sigma},
	\label{eq:ham_kin_real}
\end{eqnarray}
where $\cdag_{i\sigma}$ ($c_{i\sigma}$) creates (annihilates) an electron with spin $\sigma$ at the $i$-th lattice site. The notation $\langle \cdots \rangle$ indicates hopping is restricted to the nearest neighbor (NN), $t$ is the amplitude of the NN hopping integral, and $\mu$ is the chemical potential. To obtain the band structure, we perform a Fourier transformation and write the Hamiltonian in momentum space
\begin{eqnarray}
	\mathcal{H}_{\text{KIN}}= \sum_{\mathbf{k}\sigma} \psi^{\dagger}_{\mathbf{k}\sigma} h_{\text{KIN}}(\mathbf{k}) \psi_{\mathbf{k\sigma}}.
\end{eqnarray} 
Here the basis is given by $\psi_{\mathbf{k}\sigma} = [c_{\mathbf{k}A\sigma}, c_{\mathbf{k}B\sigma}, c_{\mathbf{k}C\sigma}]^T$, where $c_{\mathbf{k}s\sigma}$ is an annihilation operator of electron with spin $\sigma$ and momentum ${\bm k}$ in sublattice $s \in \{A,B,C\}$. 
The kernel of the Hamiltonian reads
\begin{eqnarray}
	h_{\text{KIN}}(\mathbf{k}) =
        \begin{bmatrix}
            -\mu          & -2t\cos\delta_1 & -2t\cos\delta_3 \\
            -2t\cos\delta_1 & -\mu          & -2t\cos\delta_2 \\
            -2t\cos\delta_3 & -2t\cos\delta_2 & -\mu          
        \end{bmatrix}.
        \label{eq:ham_k_kin}
\end{eqnarray}
For brevity, we use the notation $\delta_i=\mathbf{k}\cdot \mathbf{d}_i$, where $\mathbf{k}=(k_x,k_y)$ is the 2D wavevector, and $\mathbf{d} \in \{\mathbf{a}_1/2, \mathbf{a}_2/2, (\mathbf{a}_1-\mathbf{a}_2)/2 \}$ is the vector connecting NN sites [see Fig.~\ref{fig:schematic}(a)]. The eigenvalues of Eq.~(\ref{eq:ham_k_kin}) are given by  $E(\mathbf{k})=2t, -t\big[1 \pm \sqrt{f(\mathbf{k})} \big]$ with $f(\mathbf{k}) = \big[3+2\cos k_x + 4\cos(k_x/2)\cos(\sqrt{3}k_y/2) \big]$, for $\mu$ set to zero. We plot the band structure along the high symmetry path in Fig.~\ref{fig:schematic}(b), which shows two dispersive and one flat band (FB). Notably, the dispersive bands contain a Dirac point (DP) at K (K') and saddle points at M, which leads to a logarithmic van Hove singularity (VHS) in the density of states (DOS) [see right panel in Fig.~\ref{fig:schematic}(b)].

\section{Crystalline edge states: termination dependence}
\label{sec:slab_termination}

We begin by examining the edge states of the pristine kagome lattice described by Eq.~\eqref{eq:ham_kin_real}. Edge states are important because they can substantially alter the electronic properties of a system, and provide direct insight into bulk topology. Since edge states cannot exist in an infinite lattice without boundaries, their study requires an explicit lattice termination. This prompts us to consider a ``slab geometry'' to study the edge states of kagome lattice. A 2D slab structure implies that one of the directions is under periodic boundary conditions, while the other direction is finite. For simplicity, we restrict ourselves to line cuts  along the main axes in this work. This leads to four types of edge patterns on the kagome lattice: zigzag, armchair, cove, and flat [see Fig.~\ref{fig:schematic}(a)]. To showcase the periodicity of the slab, we present the Brillouin zone (BZ) in Fig.~\ref{fig:schematic}(c). The hexagon represents the bulk BZ, where the Dirac points are symmetrically located at the corners. A slab with edges at fixed $x$ has periodic BZ along $k_y$ direction --- this leads to the high symmetry path: $\mathrm{\overline{K}}$--$\mathrm{\overline{\Gamma}}$--$\mathrm{\overline{K}}$. Similar arguments can be used for other directions.

To analyze the edge states, we focus on the slab band structure obtained from exact diagonalization of Eq.~\eqref{eq:ham_kin_real}. We are also interested in how the edge states are distributed in the real-space lattice, and for this, we examine the local density of states (LDOS). Since the edge state does not always lie inside a fully gapped spectrum, we use the momentum-resolved site LDOS: 
\begin{eqnarray}
\rho_i(E, k_a) = \sum_n |\psi_{in}(k_a)|^2 \; \delta[E-E_n(k_a)],
\end{eqnarray}
Here, $\psi_{i n}$ is the wavefunction of the $n$-th state with energy $E_n(k_a)$, $a = x/y$ labels the periodic direction, and $i=1,2,\cdots,N$ refers to site index with $N$ being number of sites. We approximate the Dirac delta function by a Lorentzian $\delta(x) = (\epsilon/\pi)/(x^2+\epsilon^2)$, with a small broadening $\epsilon = 10^{-3}$. For illustrative purposes, we usually pick the energy $E$ to correspond to an edge state at momentum $k_a$ that maximizes the number of edge states relative to bulk states.

%
\begin{figure*}[t]
	\includegraphics[width=\linewidth]{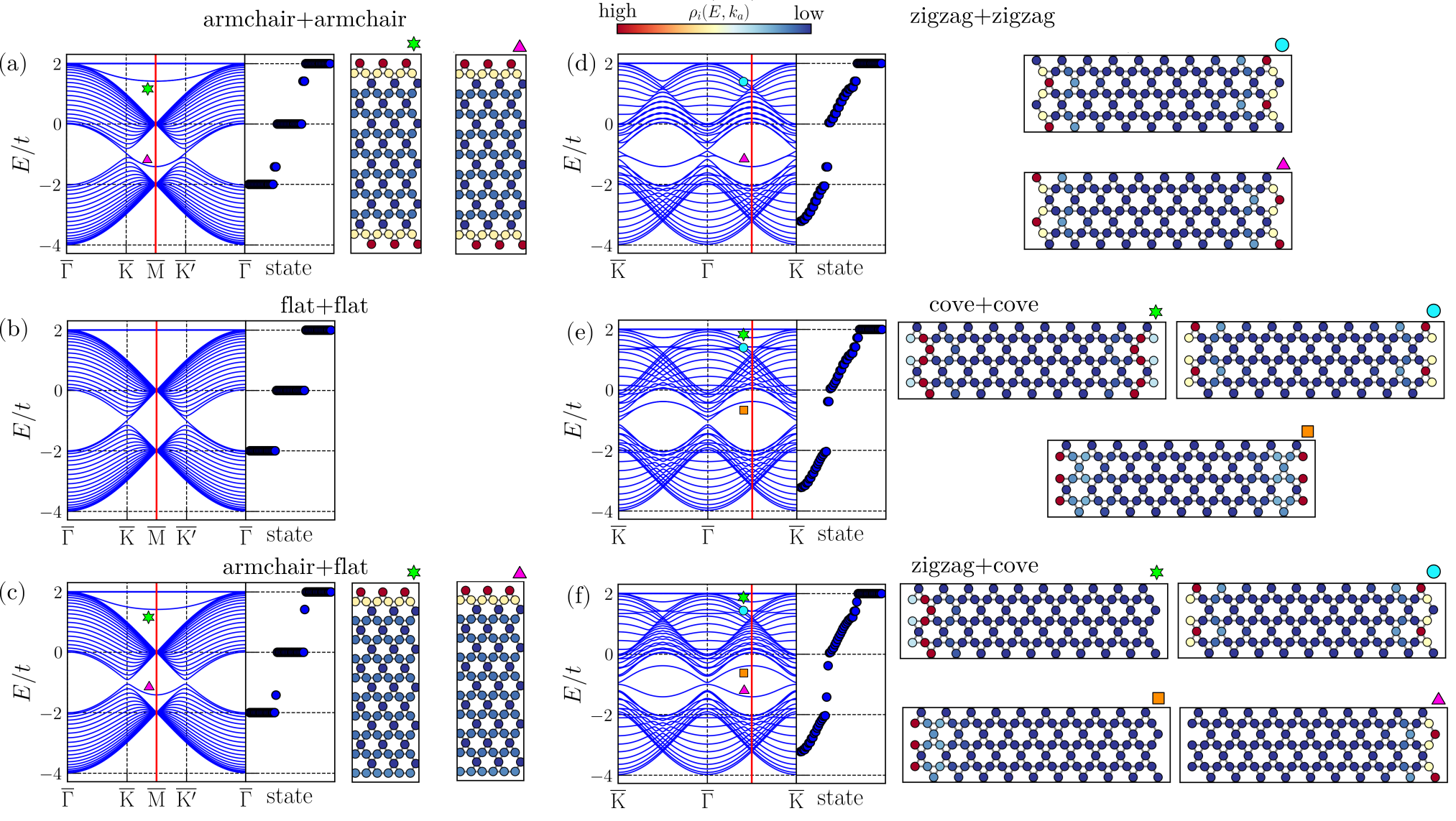}
	\caption{Kagome slab band structure for different lattice terminations (left), with energy levels at a specific momentum (red vertical line) to pinpoint the edge states (middle), edge states are highlighted by icons, and we present the momentum-resolved site LDOS on real-space lattice (right). Parameters are $\mu/t=0$ and $t=1$.}
	\label{fig:cryst-symm}
\end{figure*}
%

\subsection{Edge types at fixed $y$}

First, we describe three possible terminations at fixed $y$ axis: armchair or flat terminations for both slab edges or one for each edge. The armchair termination hosts
two edge states in the band structure [see~Fig.~\ref{fig:cryst-symm}(a)]: one that connects the DPs from $\mathrm{\overline{K}}$ and $\mathrm{\overline{K}}^{\prime}$, while the other exists near the FB, see triangle and star icons, respectively. For further analysis, we pick the $\mathrm{\overline{M}}$ point (red vertical line), and plot the energy levels against the state index, next to the band structure. The spectrum shows three continuous spectra arising from the bulk bands, plus two pairs of isolated states located in the gaps. These gapped states lie at opposite energies and are two-fold degenerate. Here, the degeneracy stems from the similarity between the top and bottom edges. It is worthwhile to mention that, even with similar edge types, mirror symmetry may not always be present. This is the case when opposite edges are connected by screw symmetry (reflection+translation), even though they are of the same type. Nonetheless, we find no difference in the band structure in the presence and absence of mirror symmetry (results not shown). Also plotting the edge state LDOS at the $\mathrm{\overline{M}}$ point, we find that the electrons are mostly localized at the boundaries. More specifically, the LDOS strongly peaks at the edge sites with two nearest neighbors. We notice that the LDOS map forms a triangular plaquette at the edge. Electrons (of $s$-type) moving in such a cycloid trajectory are predicted to carry a finite orbital angular momentum (OAM) ~\cite{bush.mertig.23}. Inside the bulk, neighboring upward and downward triangles negate the OAM, but this cancellation does not occur at the edges. The OAM arises due to translation and rotation and thus a finite group velocity $v_{\mathbf{k}} = \partial_{\mathbf{k}} E(\mathbf{k})$ of kagome edge states can be useful in orbitronics.   

For the flat edge termination, we find that the edge states completely disappear [see Fig.~\ref{fig:cryst-symm}(b)]. This type of termination instead yields a band structure similar to the bulk, with no in-gap edge states. For this reason, we omit the LDOS. 
We give a possible explanation of edge state suppression in the Sec.~\ref{sec.sm2} in the Supplemental Material (SM)~\footnote{See Supplemental Material at [URL will be inserted by publisher] for additional theoretical results.}.
Finally, we turn to the case where one side of the slab has armchair-termination, and the other flat termination [see Fig.~\ref{fig:cryst-symm}(c)]. We find that the resulting edge states are similar to the armchair termination. However, in contrast, the edge states are no longer two-fold degenerate due to the lack of mirror symmetry. The LDOS shows that both edge states are localized at the armchair termination, as expected from the results in Fig.~\ref{fig:cryst-symm}(a,b).

\subsection{Edge types at fixed $x$}

We next focus on edge terminations at fixed $x$ axis, which are zigzag, cove, and a mix of both. We begin by analyzing the band structure of the pure zigzag termination [see Fig.~\ref{fig:cryst-symm}(d)]. The bulk spectrum contains three DPs at the $\mathrm{\overline{K}}$, $\mathrm{\overline{K}}^{\prime}$, and $\overline{\mathrm{\Gamma}}$, respectively. Interestingly, the DP at the $\overline{\mathrm{\Gamma}}$ point appears due to projection of the K point onto the $\mathrm{\Gamma}$ point [see Fig.~\ref{fig:schematic}(c)]. This is in contrast to the slab BZ along the $k_x$ direction, which has only two DPs. Therefore, specific lattice terminations can dictate hybridization at the $\mathrm{\Gamma}$ point with other states (e.g. substrate/impurity), which is important for device applications. In the slab spectrum, there exists a pair of twofold-degenerate edge states: one connecting the DPs (triangle) and another near the FB (circle), reminiscent of the armchair case. This degeneracy arises due to mirror symmetry. For our analysis, we here take an energy cut (highlighted by red line) between the $\mathrm{\overline{\Gamma}}$ and $\mathrm{\overline{K}}$ points, and plot the LDOS. We see that the in-gap states are localized at the slab boundary as edge states. Depending on the edge state energy, the LDOS acquires a phase shift so that the maximum LDOS is on a site with either two or four NN sites.

\begin{figure*}[t]
	\includegraphics[width=\linewidth]{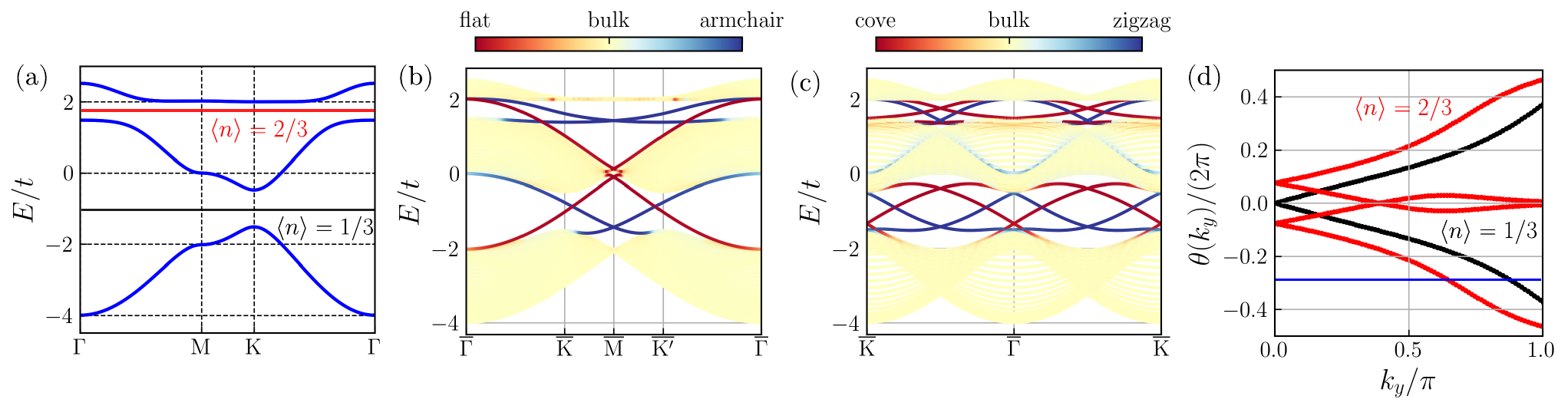}
	\caption{(a) Bulk kagome bands for the QSH phase and the slab band structure for (b) armchair-flat and (c) zigzag-cove type terminations. In all bandstructure plots, we take $\lambda_{\text{KM}}/t=0.15$ and $\mu/t=0$. Colors show expectation of the position operator. (d) Eigenvalues of the Wilson loop plotted as a function of $k_y$ for two different fillings, as indicated in (a). Eigenvalues of both fillings $\langle n \rangle$ cross the reference line (blue line) once (odd number of times), indicating the $\mathbb{Z}_2$ nature.}
	\label{fig:soc-bands}
\end{figure*}

As we move to the cove termination [see Fig.~\ref{fig:cryst-symm}(e)], a non-degenerate edge state appears in the spectrum. This state is marked by a star icon, while the others (marked by circle and square icon) are twofold degenerate due to mirror symmetry. The LDOS shows that all edge states are localized, but with some spread into the bulk. Here, only the edge state close to zero energy localizes on a dangling site; other edge states are localized on non-dangling edge sites with four NN sites, further into the slab.

Finally, we consider a mix of zigzag and cove terminations [see Fig.~\ref{fig:cryst-symm}(f)]. We find that mixing both terminations yields all possible edge states from each termination. At the same time, this breaks the mirror symmetry and removes any global degeneracy of the states. However, point degeneracies are still possible, for example, at the momentum marked by a red vertical line. We thus find five non-degenerate edge states in total. Notably, the LDOS shows that only the edge state marked by the circle is localized on both sides of the slab. In all other cases, the edge states are localized on the cove termination. 
Overall, we find that the pristine kagome lattice hosts a multitude of edge states, strongly dependent on the lattice terminations.

\section{Quantum spin Hall phase and $\mathbb{Z}_2$ classification}
\label{sec:soc}

Kagome materials are known to exhibit intrinsic SOC. It was first reported in the ferromagnetic nodal-line semimetal $\mathrm{Fe_3Sn_2}$~\cite{ye.chan.19,fang.ye.22}, where SOC fully gaps out the nodal lines, leading to massive Dirac bands. Later, intrinsic SOC was also confirmed in 135~\cite{ortiz.teicher.20} and 166~\cite{sante.bigi.23} kagome materials.
In this section, we investigate how the crystalline edge states change under intrinsic SOC. More specifically, we choose  Kane--Mele type SOC (KMSOC), which is a prototypical model for the quantum spin Hall (QSH) effect. The SOC Hamiltonian can be written as
\begin{eqnarray}
H_{\text{QSH}} &=& H_{\text{KIN}} + H_{\text{KM}},
\end{eqnarray}
Here, the first term refers to the kinetic part described in Eq.~\eqref{eq:ham_kin_real}, while the latter term denotes the KM part
\begin{eqnarray} 
\label{eq:ham_km_soc} 
H_{\text{KM}} &=& i\lambda_{\text{KM}} \sum_{\langle\langle ij \rangle\rangle \sigma} \nu_{ij} c^{\dagger}_{i\sigma} \sigma_z c_{j\sigma},
\end{eqnarray}
where $\lambda_{\text{KM}}$ is the strength of the coupling and $\sigma_z$ is the third Pauli matrix acting in spin subspace. The notation $\langle\langle \cdots \rangle\rangle$ means KMSOC is a NNN hopping with a chiral factor $\nu$, which takes values $1$ or $-1$ if the hopping is anticlockwise or clockwise, respectively. Note that $H_{\text{KM}}$ preserves time-reversal symmetry (TRS) since opposite spins are assigned opposite chiralities.

First, we look at the bulk bands to get insight into the KM term. In the momentum space, we write the KM Hamiltonian in the three-sublattice ($A,B,C$) kagome spinful ($\uparrow, \downarrow$) basis $\psi_{\mathbf{k}} = [c_{\mathbf{k}A\uparrow}, c_{\mathbf{k}B\uparrow}, c_{\mathbf{k}C\uparrow}, c_{\mathbf{k}A\downarrow}, c_{\mathbf{k}B\downarrow}, c_{\mathbf{k}C\downarrow}]^T$ 
\begin{align} \label{eq:ham_qsh}
h_{\text{KM}}(\mathbf{k}) &=  
\sigma_z \otimes 2i\lambda_{\text{KM}} 
\begin{bmatrix}
0 & \cos\tau_3 & -\cos\tau_1 \\
-\cos\tau_3 & 0 & \cos\tau_2 \\
\cos\tau_1 & -\cos\tau_2 & 0
\end{bmatrix}, 
\end{align}
where $\tau_i = \mathbf{k}\cdot \mathbf{d}^{\prime}_i$, $\mathbf{k}$ is the wavevector, and $\mathbf{d}^{\prime}_i$ are the vectors connecting NNN sites.

The bulk band structure is shown in Fig.~\ref{fig:soc-bands}(a). A finite KM coupling $\lambda_{\textrm{KM}}/t=0.15$ opens band gaps at high-symmetry points where bands were originally touching: one at the K point between the first and second dispersive bands (counted from the bottom), and another at the $\Gamma$ point between the second dispersive and third flat band. Furthermore, the presence of KMSOC renders the flat band dispersive.

The opening of band gap leads to gapless edge states in a slab geometry [see Fig.~\ref{fig:soc-bands}(b,c)]. Here, we consider two types of slabs with mixed edges, and keep the slab width large enough to avoid any hybridization between edge states. It is also judicious to project the expectation of the position operator on the bands to pinpoint edge state localization~\cite{sekh.chakrabarti.20}. For this, we introduce the position-weighted amplitude
\begin{eqnarray}
\rho_{n,a} (\mathbf{k}) = \sum_i |\psi_{in}(\mathbf{k})|^2 \; \mathbf{r} \cdot \hat{e}_{a},
\end{eqnarray} 
where $\mathbf{r}$ is the position operator, $\psi_{in}$ is the wavefunction of the $n$-th state, and $\hat{e}_{a}$ is the unit vector along $a$-direction.
For both slab directions, we find that each boundary contains two localized edge states in each gap [see Fig.~\ref{fig:soc-bands}(b,c)]. What is interesting is that the SOC enables the flat boundary to host edge states, even though it does not exhibit any edge states without KMSOC. This termination-independence hints at topological edge states, which depend on bulk topology rather than boundary details. For the zigzag-cove edge type, we also notice that bands at $\mathrm{\overline{\Gamma}}$  and $\mathrm{\overline{K}}$ are similar -- this happens due to projection of the DP on $\mathrm{\overline{\Gamma}}$ as shown in Fig.~\ref{fig:schematic}(c). We can verify the presence of TRS since $E(\mathrm{\overline{K}})=E(\mathrm{\overline{K}}^{\prime})$.

To classify the topology of the TRS bands, we compute the $\mathbb{Z}_2$ index, which is the topological invariant for QSH~\cite{kane.mele.05, kane.mele.z2.05} phase. A time-reversal invariant QSH phase has $\mathbb{Z}_2=1$, which distinguishes it from ordinary insulators with $\mathbb{Z}_2=0$. One of the ways to compute the $\mathbb{Z}_2$ index is through the Wilson loop formalism~\cite{yu.qi.11,alex.dai.14}. This method is equivalent to tracking maximally localized Wannier centers~\cite{soluyanov.vanderbilt.11.1,soluyanov.vanderbilt.11.2}. In this approach, we define the transfer matrix in the discretized BZ
\begin{eqnarray}
W(k_y) = \prod_i \; \mathbf{P}_{i,i+1} (k_y),
\end{eqnarray}
where $P^{mn}_{i,i+1} (k_y) = \langle u_m(k^i_x, k_y) | u_n(k^{i+1}_x, k_y) \rangle$ is the overlap matrix, $(m,n)$ refer to occupied bands, and $|u_n\rangle$ is the Bloch wavefunction. Given $N_{\text{occ}}$ occupied bands, $W(k_y)$ is a $N_{\text{occ}}\times N_{\text{occ}}$ matrix with eigenvalues $\lambda_1, \lambda_2, ..., \lambda_{N_{\text{occ}}}$. Then the phase of the eigenvalues is
\begin{eqnarray}
\theta_n (k_y) = \textrm{Im} [\ln\lambda_n (k_y)].
\end{eqnarray}
The winding of the phase factor $\theta(k_y)$ for each eigenvalue against $k_y$ is key to infer the $\mathbb{Z}_2$ index. More specifically, if $\theta(k_y)$ crosses a reference line $\theta_c$ odd number of times, we have $\mathbb{Z}_2=1$, otherwise it is trivial and $\mathbb{Z}_2=0$.

\begin{figure*}[t]
	\includegraphics[width=\linewidth]{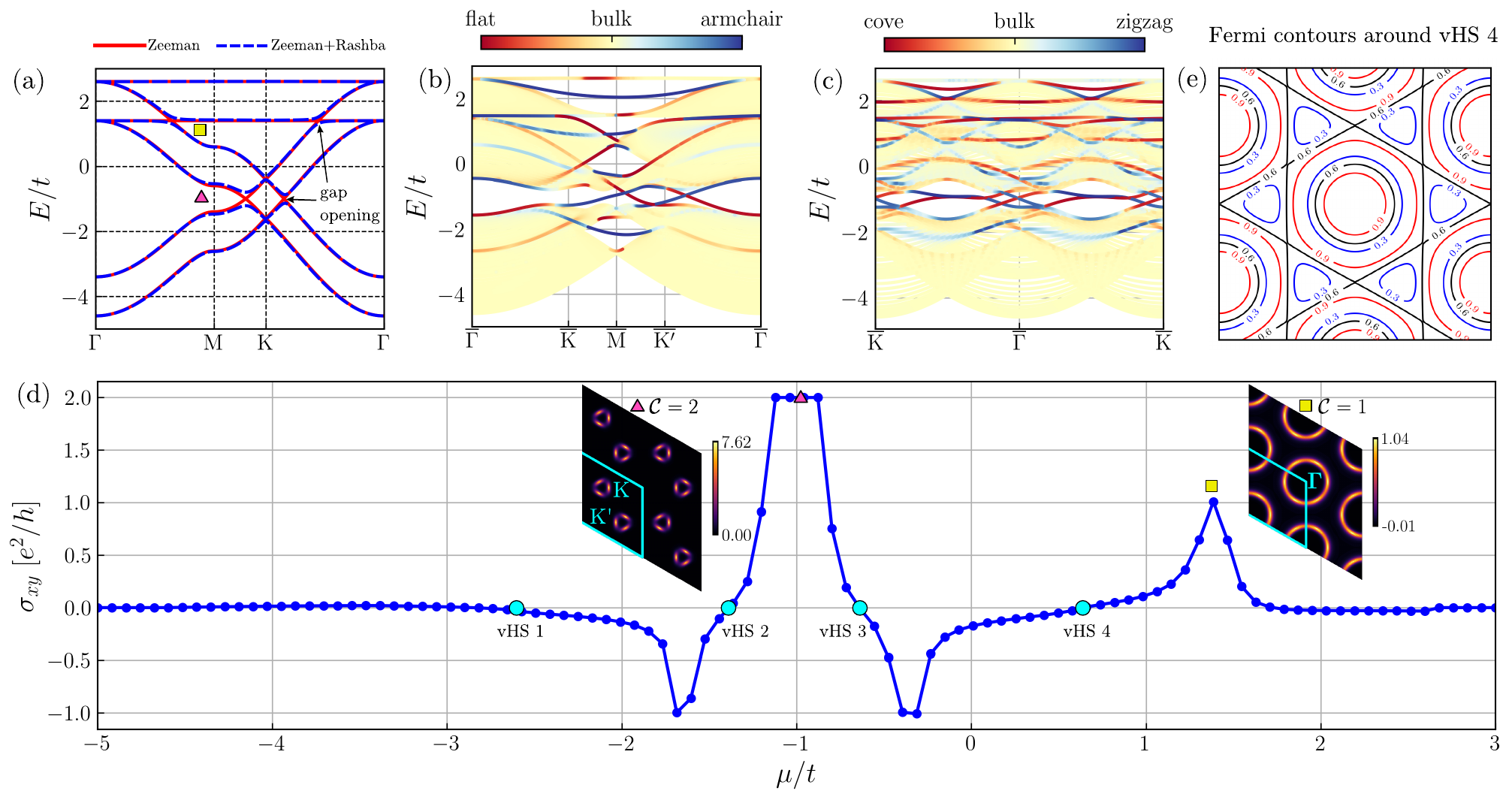}
	\caption{(a) Bulk kagome bands for the QAH phase without (red) and with RSOC ($\lambda_{R}/t=0.2$) (blue dashed) under constant Zeeman field ($h_z/t=0.6$). Slab band structures with Zeeman and RSOC for (b) armchair-flat and (c) zigzag-cove terminations,  with colors showing the expectation value of the position operator. (d) Anomalous Hall conductivity (in units of $e^2/h$) as a function of chemical potential. Circular cyan icons highlight the location of the VHS points. Insets show the Berry curvature maps in the two topological gaps, with icons also indicated in (a). Enclosed cyan line denote the BZ. (e) Fermi contours plotted at ($\mu/t=0.6$) and around $\mu/t=(0.3,0.9$) the VHS 4 in (d).}
	\label{fig:ahe_wo_km}
\end{figure*}

We show the evolution of $\theta(k_y)$ in Fig.~\ref{fig:soc-bands}(d) at two different fillings. We recall that the number of eigenvalues is determined by the number of occupied bands. Owing to spin degeneracy, there are two occupied bands at $1/3$ filling and four at $2/3$ filling, which accounts for the corresponding number of solutions at each filling. Due to TRS, $k_y$ is evolved through half of BZ; in fact, $k_y=0$ and $k_y=2\pi$ are Kramers-degenerate points related by TRS. As $k_y$ is varied from $0$ to $\pi$, we observe that $\theta(k_y)$ exhibits an approximately $\pi$ jump for both fillings. Importantly, upon choosing a reference line $\theta_c=-0.3$, $\theta(k_y)$ crosses this line once (or, more generally, an odd number of times), signaling non-trivial $\mathbb{Z}_2$ bulk topology.

A direct consequence of the $\mathbb{Z}_2$ phase is the presence of helical edge states, namely, pairs of spin-polarized edge modes that propagate in opposite directions. This is thus the origin of edge states in Fig.~\ref{fig:soc-bands}(b,c) in both bulk gaps, where the edge states associated with a given boundary exhibit opposite slopes, indicating counter-propagating motion. Furthermore, these counter-propagating modes carry opposite spin polarizations, as confirmed by the spin-polarized band structure in Fig.~\ref{fig.sm:sp-ldos} in the SM~\cite{Note1}.
Thus, the inclusion of KMSOC gaps out the otherwise gapless edge modes of the pristine kagome lattice, yielding helical, spin-polarized edge states that are consistent with a $\mathbb{Z}_2$ topological phase. The resulting edge dispersion respects the $\mathbb{Z}_2$ symmetry, such that the edge modes appear in Kramers pairs, are spin-polarized, and their number is independent of the lattice termination.

\section{Quantum anomalous Hall phase via ferromagnetism}
\label{sec:qah}
\subsection{Zeeman field and Rashba SOC}

The QSH phase discussed in Sec.~\ref{sec:soc} evolves into a QAH phase once TRS is broken~\cite{yang.xu.11}. The key signature of this phase is an anomalous Hall conductivity, which is a Hall response generated without any magnetic field~\cite{chang.zhang.13}. Motivated by the occurrence of ferromagnetism (FM) in kagome materials~\cite{ye.kang.18,liu.sun.18,chen.le.21}, we use a Zeeman field to provide a simple mechanism to break TRS. In fact, the kagome lattice generally supports FM due to the presence of a non-dispersive FB~\cite{mielke.99} or a VHS~\cite{stoner.38}, if appearing near the Fermi level. In magnetic kagome semimetals, the inclusion of SOC lifts the nodal line degeneracies~\cite{kuroda.tomita.17,liu.liu.22}, which can be modeled by RSOC~\cite{sekh.mandal.22}.

The anomalous Hall conductivity is known to be quantized~\cite{thouless.kohmoto.82,haldane.88,yu.zang.10} in units of $e^2/h$ inside a global topological gap. This is called a Chern insulating phase, which is characterized by a $\mathbb{Z}$ invariant known as the Chern number.
To achieve a quantized QAH, pairing the Zeeman, or exchange, field with Rashba SOC (RSOC) is often useful. While the exchange field creates spin-polarized bands, the RSOC can further lift spin degeneracy in momentum space to create a bulk gap~\cite{qiao.yang.10,ding.qiao.11}. The RSOC leads to so-called ``spin-momentum locking'' -- a phenomenon where spin orientation depends on momentum, and as a result, the bulk bands exhibit a momentum-dependent splitting. An important distinction is that, unlike KMSOC, the RSOC mixes the spin sector, so that $S_z$ is no longer a conserved quantum number. The appearance of a Rashba term requires broken inversion symmetry, which in practice can arise due to non-centrosymmetric stacking of layers or surface/interface effects.

We add FM and RSOC in the kagome lattice through the following minimal Hamiltonian
\begin{eqnarray}
H_{\text{QAH}} = H_{\textrm{KIN}} + H_{\textrm{R}} + H_{\textrm{Z}},
\end{eqnarray}
where the first term is introduced in Eq.~\eqref{eq:ham_kin_real}. The second and third terms describe the Hamiltonian for RSOC and out-of-plane Zeeman field, respectively
\begin{eqnarray}
H_{\textrm{R}} &=& - i \lambda_R \sum_{\langle ij \rangle,\sigma\sigma^{\prime}} c^{\dagger}_{i\sigma} (\mathbf{d}_{ij} \times \boldsymbol{\sigma}_{\sigma\sigma^{\prime}})_z \, c_{j\sigma^{\prime}}, \\
H_{\textrm{Z}} &=& -h_z \sum_i (c^{\dagger}_{i\uparrow}c_{i\uparrow} - c^{\dagger}_{i\downarrow}c_{i\downarrow}).
\end{eqnarray}
In the RSOC Hamiltonian, $\lambda_R$ denotes the strength of the spin-orbit coupling. The Zeeman field creates a splitting of $h_z$ between opposite spins at the $i$-th site.

In momentum space, we express the Hamiltonian in the same basis as $h_{\textrm{KM}}$ in Eq.~\eqref{eq:ham_qsh}:
\begin{eqnarray}
h_{\textrm{R}} (\mathbf{k}) &=& \sigma_y \otimes \lambda_R 
	\begin{bmatrix}
		0          & \sin\delta_1 & \gamma^{*}\sin\delta_3 \\
		\sin\delta_1 & 0          & \gamma\sin\delta_2  \\
		\gamma^{*}\sin\delta_3  & \gamma\sin\delta_2 & 0
	\end{bmatrix}, \nonumber \\
h_{\textrm{Z}} (\mathbf{k}) &=& \sigma_z \otimes 
	\begin{bmatrix}
		h_z          & 0 & 0 \\
		0 & h_z          & 0  \\
		0  & 0 & h_z
	\end{bmatrix},
\end{eqnarray}
where we denote $\delta_i=\mathbf{k}\cdot \mathbf{d}_i$ and $\gamma=i\exp(i\pi/6)$. We illustrate the bulk band structure in Fig.~\ref{fig:ahe_wo_km}(a). The Zeeman field leads to spin-polarized and -split bands, which create a pair of DP and FB, as well as four VHS points. Applying RSOC mixes the spin channels and opens up a gap at the DP and FB. These bulk gaps host gapless edge states, which can be seen in the slab band structure in Fig.~\ref{fig:ahe_wo_km}(b,c). Specifically, two pairs of edge states exist in the global gap around DP, whereas the pair of edge states near FB are less visible due to the small band gap. 

\begin{figure*}[t]
	\includegraphics[width=\linewidth]{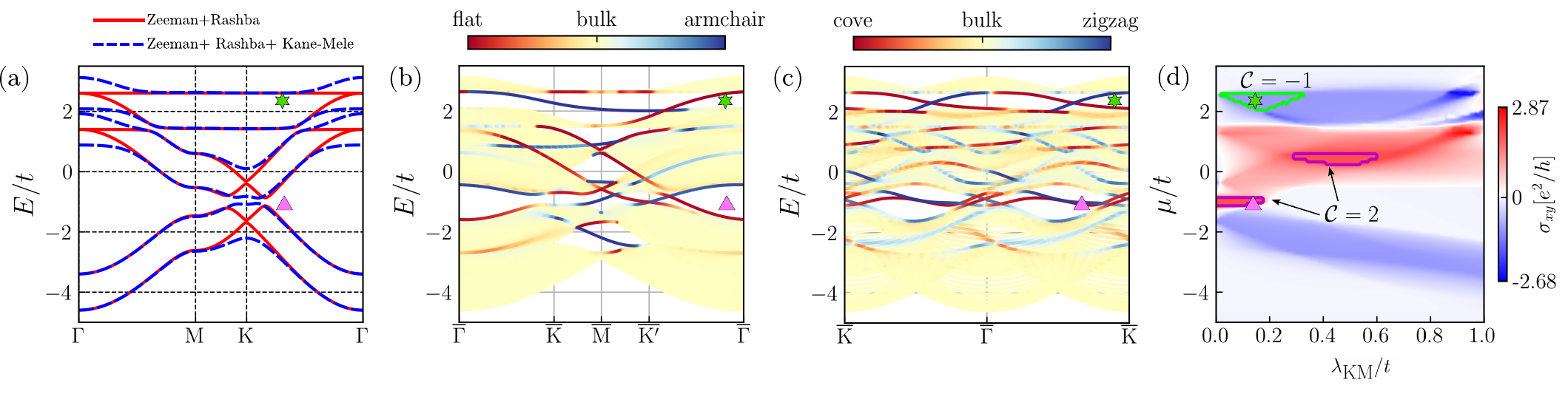}
	\caption{(a) Bulk kagome bands for the QAH phase without (red) and with KMSOC ($\lambda_{\textrm{KM}}/t=0.15$) (blue dashed) under constant Rashba SOC ($\lambda_{R}/t=0.2$) and Zeeman field ($h_z/t=0.6$). Slab band structures with Zeeman, RSOC, and KMSOC for (b) armchair-flat and (c) zigzag-cove terminations, with colors showing the expectation of the position operator. (d) Phase diagram capturing the evolution of the anomalous Hall conductivity $\sigma_{xy}$ (in terms of $e^2/h$) as a function of chemical potential $\mu$ and Kane--Mele coupling $\lambda_{\mathrm{KM}}$. Enclosed regions indicate quantized QAH phases with finite Chern number $\mathcal{C}$. In all plots, triangle and star icons refer to different topological gaps with $\mathcal{C}=2$ and $\mathcal{C}=-1$, respectively.}
	\label{fig:ahe_w_km}
\end{figure*}

The opening of a bulk gap and existence of edge states cannot verify non-trivial topology alone. We therefore also compute the intrinsic anomalous Hall conductance~\cite{jungwirth.niu.02,nagaosa.sinova.10}
\begin{eqnarray} \label{eq:ahe_cond}
\sigma_{xy} (E) = \frac{e^2}{h} \int_{\textrm{BZ}} \frac{d^2k}{A_{\textrm{BZ}}} \sum_n \Omega_{n,\mathbf{k}} \; \Theta (E_{n,\mathbf{k}} - E),
\end{eqnarray}  
where $\Theta$ denotes the unit step function, $E_n$ is the energy of the $n$-th band, $A_{\textrm{BZ}}$ stands for the area of the BZ, and the Berry curvature is given by the formula 
\begin{eqnarray}
\Omega_{n,\mathbf{k}} &=& - \mathrm{Im} \sum_{m\neq n} \frac{F_{mn,\mathbf{k}}^{xy} - F_{mn,\mathbf{k}}^{yx}}{(E_n - E_m)^2},
\\ \nonumber
F_{mn, \mathbf{k}}^{xy} &=& \langle u_n | \partial_{k_x} H_{\text{QAH}}  | u_m \rangle \langle u_m | \partial_{k_y} H_{\text{QAH}} | u_n \rangle.
\end{eqnarray}
Breaking of TRS implies $\Omega_{n,\mathbf{k}}\neq \Omega_{n,-\mathbf{k}}$, which renders the anomalous conductance nonzero in Eq.~\eqref{eq:ahe_cond} -- a hallmark of QAH, where the Hall response is generated without the magnetic field. We plot $\sigma_{xy}$ as a function of $\mu/t$ in Fig.~\ref{fig:ahe_wo_km}(d). The conductance becomes finite as more bands are filled, and exhibits a perfect quantized plateau around $\mu/t=-1$. This corresponds to the bulk gap around the DP [see pink triangle in Fig.~\ref{fig:ahe_wo_km}(a,d)]. Inside a global topological QAH gap, the anomalous conductance is a constant given by $\sigma_{xy} = \mathcal{C} e^2/h$, where $\mathcal{C}$ is the Chern number of the topological gap. Here we find $\mathcal{C}=2$ for the gap around DP.
Due to the bulk-boundary correspondence, this implies two protected edge states per edge, which can be seen in Fig.~\ref{fig:ahe_wo_km}(b,c). Similarly, the narrow gap near the flat band [see yellow square in Fig.~\ref{fig:ahe_wo_km}(a,d)] is topological with $\mathcal{C}=1$, although the edge states are poorly resolved due to the small gap.
Note that for the zigzag-cove termination, the BZ has a period given by $\overline{\Gamma}$ and $\overline{\mathrm{K}}$, such that Fig.~\ref{fig:ahe_wo_km}(c) displays the chiral modes twice. Also note that according to Fig.~\ref{fig:cryst-symm}, the armchair-flat terminations are the top-bottom edges, while the cove-zigzag terminations are the left-right edges, respectively, which set how the Chern number sign determines the edge state dispersion. 

The insets of Fig.~\ref{fig:ahe_wo_km}(d) show the occupied Berry curvature at the two topological gaps. The map for $\mathcal{C}=2$ contains satellite-like Berry curvature, with peaks concentrated near K and K$^{\prime}$. Here, the lack of TRS ensures the satellite peaks are asymmetric between valleys. A less common pattern occurs for $\mathcal{C}=-1$, where the Berry curvature peaks form a ring around $\Gamma$. Other than these two global gaps, $\sigma_{xy}$ sometimes shows a nearly quantized behavior. This is, however, not related to a finite Chern number since the system is here metallic, which makes the Chern number ill-defined.

Finally, we dedicate a few words on the relation between the change of sign in $\sigma_{xy}$ and the Fermi surface (FS) topology. In Fig.~\ref{fig:ahe_wo_km}(d), $\sigma_{xy}$ changes sign at four points (cyan circles), which correspond to the VHS saddle points of the band structure. At these VHS, the FS abruptly changes its topology without necessarily breaking any symmetry. This is known as a Lifshitz transition~\cite {lifshitz.60,volovik.17}. To illustrate this effect, we show the FS contour around VHS 4 in Fig.~\ref{fig:ahe_wo_km}(e). The FS at the VHS reflects the lattice symmetry, forming hexagonal contours of maximal length that produce a diverging DOS. Below this filling ($\mu/t=0.3$), we find a circular FSs around $\Gamma$ and K/K$^\prime$. One of these pockets disappear as the Fermi level is instead tuned above the VHS ($\mu/t=0.9$). As a result, $\sigma_{xy}$ can undergo a sign change at the VHS. Although we show the FS around VHS 4 here, the effect is similar at other VHS points.

\subsection{Impact of Kane--Mele SOC}

So far, we discussed how the RSOC opens topological Chern gaps in the spectrum. Here, we investigate how the Kane--Mele term modifies this topological landscape. If we focus on the bulk bands, we find in Fig.~\ref{fig:ahe_w_km}(a) that a finite KMSOC ($\lambda_{\mathrm{KM}}/t=0.15$) lifts point degeneracies at high symmetry points. The resulting gaps are different from those observed in the absence of KMSOC in Fig.~\ref{fig:ahe_wo_km}(a). For example, a more clear gap opens up between the FB at $\mathrm{\Gamma}$ [see green star], while the gap at K also changes [see magneta triangle]. These gaps are topological in nature. In fact, the Chern number for the first (triangle) gap is $\mathcal{C}=2$, while for the latter (star) is $\mathcal{C}=-1$. This is in agreement with the slab band structure in Fig.~\ref{fig:ahe_w_km}(b,c), where each gap hosts either one or two edge states per edge channel, consistent with the Chern number, with the Chern number sign determining the edge state dispersion.

Finally, we show the phase diagram in Fig.~\ref{fig:ahe_w_km}(d), by plotting the Hall conductance as a function of filling and Kane--Mele SOC strength. We identify quantization by the condition: $|\sigma_{xy}-\mathrm{round}(\sigma_{xy})| < \epsilon$, where $\epsilon=10^{-8}$ is a small number. The quantized regions of the phase diagram are marked with closed contours. For smaller KMSOC, both gaps near DP and FB remain topological with quantized anomalous conductance. As $\lambda_{\mathrm{KM}}/t$ is increased, the gaps around these energies become trivial, while instead a new quantized region appears close to $\mu/t=0$.

\begin{figure}[t]
	\includegraphics[width=\linewidth]{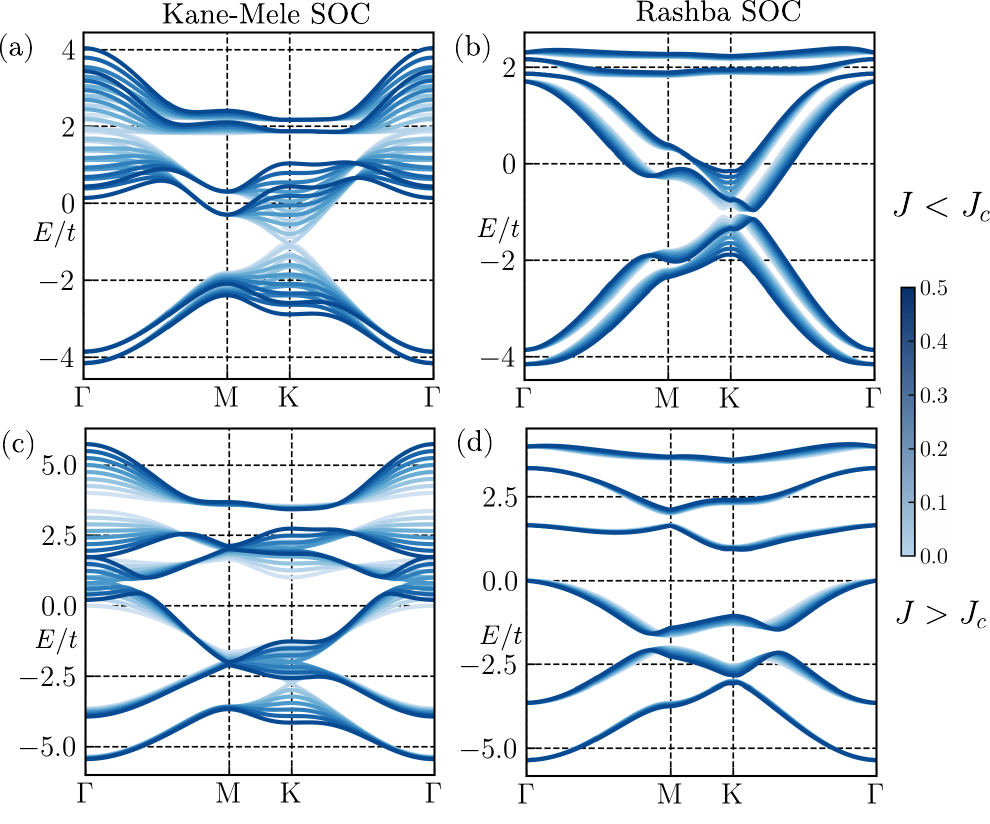}
	\caption{Effect of (a,c) Kane--Mele and (b,d) Rashba SOC going from 0 to 0.5 (blue color) on the kagome bulk bands with non-coplanar $\mathbf{q}=0$ magnetic order. Here, we consider two types of moments: $J=0.3t<J_c$ (top panel) and $J=2t>J_c$(bottom panel). Parameters are $\mu/t=0, \theta=\pi/3$, and $J_c/t \approx 1.51$.}
	\label{fig:ncl_bulk_soc}
\end{figure}

\begin{figure*}[t]
	\includegraphics[width=\linewidth]{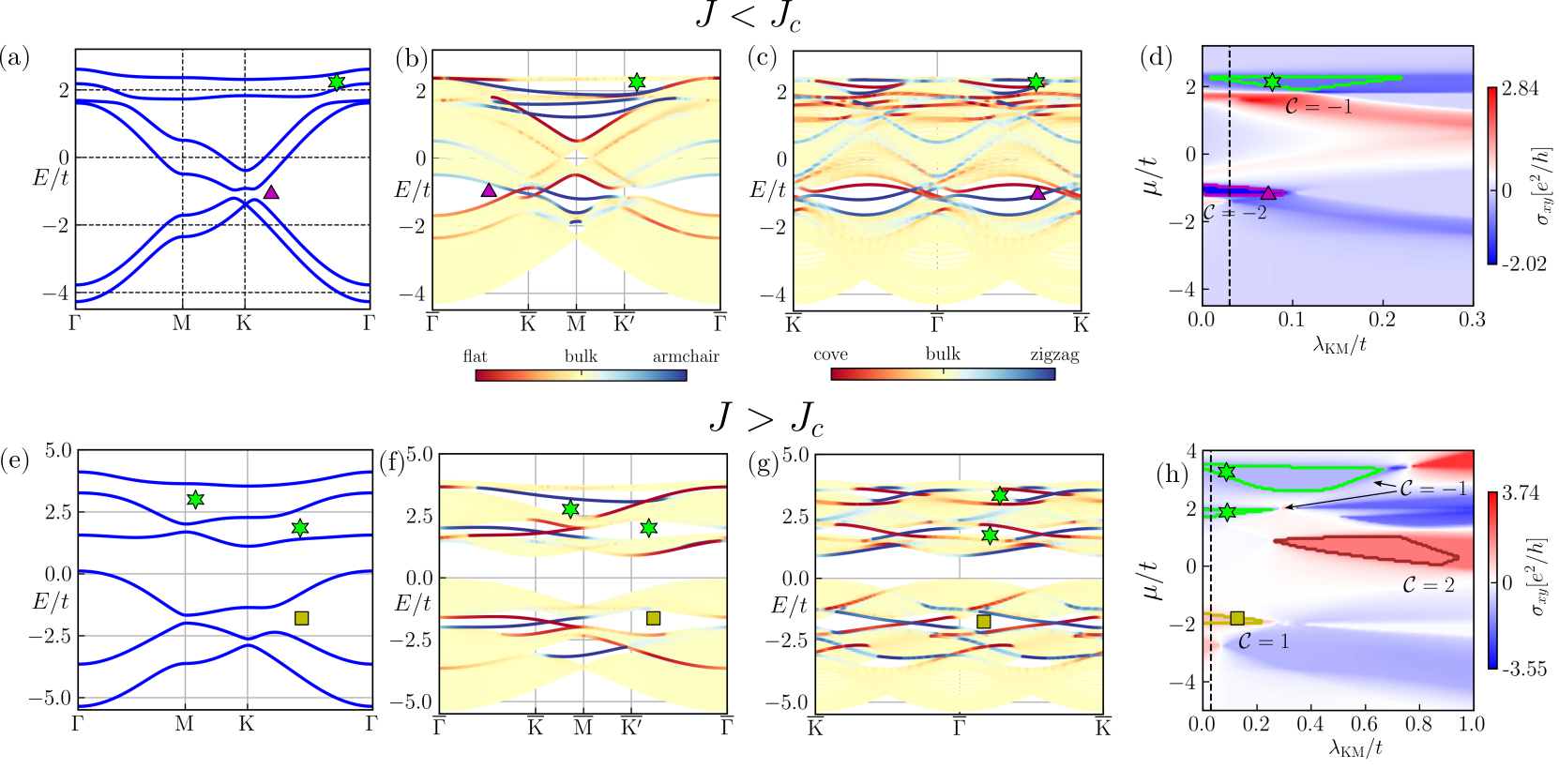}
	\caption{Kagome bulk (a,e) and slab (b,c,f,g) band structure along with topological phase diagram (d,h) for subcritical $J/t=0.5$ (top) and supercritical $J/t=2$ (bottom) exchange fields. Colors in the slab band structure shows the expectation value of the position operator. Phase diagram shows the anomalous Hall conductivity $\sigma_{xy}$ (in units of $e^2/h$) as a function of chemical potential $\mu$ and Kane--Mele coupling $\lambda_{\mathrm{KM}}$. 
   Enclosed regions indicate quantized QAH phases with finite Chern number $\mathcal{C}$. In all plots triangle, star, and square icons refer to different topological gaps with $\mathcal{C}=-2$, $\mathcal{C}=-1$, and $\mathcal{C}=1$, respectively. Horizontal dashed line marks $\lambda_{\mathrm{KM}}/t=0.03$ used in the band-structure plots. Parameters are $\mu/t=0$, $\theta=\pi/3$, and $J_c(\theta)/t\approx 1.51$.}
	\label{fig:ncl_edge_states}
\end{figure*}

\section{Non-coplanar magnetism}
\label{sec:ncl}
Finally, we turn to the case where the QAH is driven by magnetic order other than FM. Due to the frustrated geometry of the kagome lattice, several exotic magnetic orderings are possible. The simplest scenario occurs when spins on three sublattices make 120$^\circ$ angle with respect to each other, forming a magnetic structure known as the $\mathbf{q}=0$  order~\cite{chubukov.92,harris.kallin.92,gotze.farnell.11,watanabe.araki.22}. This is one of the allowed ground states of the antiferromagnetic Heisenberg model~\cite{chubukov.92,harris.kallin.92,gotze.farnell.11}, where the magnetic unit cell coincides with the lattice unit cell. Although this magnetic order is coplanar, in real materials, Dzyaloshinskii-Moriya interaction~\cite{elhajal.canals.02,li.sandhoefner.20} can lead to out-of-plane spin canting. In fact, kagome minerals~\cite{yildrim.harris.06,verrier.bert.20} and thin films~\cite{fujiwara.kato.21} are known to exhibit similar canting. Out-of-plane canting of spins forms a non-coplanar (NCL) magnetic order, which has direct consequences for the QAH phase. Particularly, scalar spin chirality~\cite{tatara.kawamura.02}, $\chi_{ijk} = \mathbf{S}_i \cdot (\mathbf{S}_j \times \mathbf{S}_k)$ ($i,j,k\in$ NN sites), a quantity closely related to TRS breaking~\cite{wen.wilczek.89}, becomes finite for NCL ordering. In contrast, a coplanar configuration obtains $\chi_{ijk}=0$. This makes the NCL magnetic order useful for realizing a QAH phase. 
In fact, previous works have predicted a QAH effect for NCL order within a Kondo-lattice model, where conduction electrons are coupled to localized spins via local exchange. The weak-~\cite{tatara.kawamura.02} and strong-coupling~\cite{ohgushi.murakami.00} regimes can be solved analytically, but the intermediate regime~\cite{taillefumier.canals.06} requires numerical solution. Despite these efforts, edge dispersion and SOC effects have been mostly neglected, which we address in this section.

To study the effects of a NCL magnetic structure, we consider the minimal local exchange Hamiltonian
\begin{eqnarray}
H_{\textrm{NCL}} = -J_0 \sum_{i\sigma\sigma^{'}} c^{\dagger}_{i\sigma} (\boldsymbol{\sigma}_{\sigma\sigma^{'}} \cdot \mathbf{S}_i) c_{i\sigma^{'}}, 
\end{eqnarray}
where magnetic moments are defined in spherical coordinates as
\begin{eqnarray}
\mathbf{S}_i = S \, (\sin\theta\cos\phi_i, \sin\theta\sin\phi_i, \cos\theta).
\end{eqnarray}
Here, $\phi_A = 0, \phi_B = 2\pi/3, \phi_C = -2\pi/3$ and $\theta$ denotes the polar angle. Furthermore, we define $J=J_0 S$ as the strength of the local exchange field. 
The bulk bands remain mostly gapless for small values of $J/t$; however, larger values typically lift pointlike degeneracies and open gaps. Such gaps are topological in nature, which means the system can be turned into a Chern insulator with Fermi level tuning. Additionally, the Chern number of bands has been found to change when $J$ crosses a critical value $J_c$, given by~\cite{taillefumier.canals.06}  
\begin{eqnarray} \label{eq:Jc}
J_c(\theta)/t = \pm \frac{2}{\sqrt{1+3\cos^2\theta}}.
\end{eqnarray} 
For our purposes, we choose the canting angle $\eta = \pi/6$, as observed in some kagome minerals~\cite{laurell.fiete.18}, which sets $\theta =\pi/2 - \eta = \pi/3$. Thus, the limit $\theta = \pi/2$ (or $\eta = 0$) corresponds to the uncanted, coplanar structure. Plugging $\theta=\pi/3$ into Eq.~\eqref{eq:Jc} yields a critical value $J_c(\theta)/t \approx 1.51$ and we expect two sets of Chern phases in the subcritical ($J<J_c$) and supercritical ($J>J_c$) regimes. 

The topological landscape is well understood without SOC, however, little is known when SOC is included. This motivates us to investigate SOC effects.
To this end, we present the evolution of the bulk band structure with KMSOC and RSOC under a constant exchange field in Fig.~\ref{fig:ncl_bulk_soc}. We consider two regimes of $J$: one below $J_c$ ($J/t=0.5$) and another above $J_c$ ($J/t=2$). Small $J/t$ is relevant in Fe-based~\cite{zhang.asmara.24} kagome materials with Hund's exchange $J_H/t \approx$ $0.25$–$0.37$, whereas kagome heavy-fermion~\cite{song.xie.25} systems with larger moments and $f$ electrons near Fermi level, are expected to show larger $J/t$. We find that tuning KMSOC in the presence of local moment significantly modifies the bands, which close/open gaps at different fillings. This behavior is present across all scales of $J/t$. In contrast, RSOC does not tune the bands nearly as much, and the energy gaps remain intact. This rules out any substantial role of RSOC in topological transitions, and we thus only focus on KMSOC.   

In the subcritical regime, the KMSOC opens multiple global gaps [see Fig.~\ref{fig:ncl_edge_states}(a)]. These gaps are then spin-split due to the exchange field. Similar to before, topological bulk gaps guarantee edge states for every termination [see Fig.~\ref{fig:ncl_edge_states}(b,c)]. Due to the size of the bulk gap, the edge states are more prominent near the DP, while the edge modes at higher energy, near the FB, hybridize with the bulk. 
To probe the non-triviality of the gaps, we plot the anomalous Hall conductivity as a function of filling and KMSOC in Fig.~\ref{fig:ncl_edge_states}(d). We find two Chern insulating regions with $\mathcal{C}=-2$ (magenta triangle) and $\mathcal{C}=-1$ (green star), enclosed by magenta and green contours, respectively. Interestingly, the $\mathcal{C}=-2$ phase can occur without KMSOC, but a finite KMSOC is required for the $\mathcal{C}=-1$ phase that sits at higher energies. Increasing $\lambda_{\textrm{KM}}/t$ transforms the chiral modes into helical-like modes, resulting in a vanishing Chern number (see Fig.~\ref{fig.sm:qah2qsh} in the SM~\cite{Note1}). An important difference between the Zeeman field and non-coplanar order is that the resulting Chern insulating phase at $1/3$ filling differs in sign, albeit with same magnitude. Note that we probe the phase diagram only up to $\lambda_{\textrm{KM}}/t=0.3$ as no quantization is seen at higher values. This is commensurate with the scale of $J/t$. 

For supercritical exchange fields, bands are mostly decoupled due to large splitting. Particularly, combination of exchange field and KMSOC leads to multiple global bulk gaps [see Fig.~\ref{fig:ncl_edge_states}(e)]. The largest gap centered around zero energy is topologically trivial, whereas the remaining gaps are topological and are indicated by icons. Specifically, yellow square and green star symbols denote Chern phases with $\mathcal{C}=1$ and $\mathcal{C}=-1$, respectively [see Fig.~\ref{fig:ncl_edge_states}(h)]. Same is corroborated by the slab band structures: each topological gap hosts a single co-propagating chiral edge mode, with its slope determined by $\text{sgn}(\mathcal{C})$ [see Fig.~\ref{fig:ncl_edge_states}(f,g)]. 
Finally, the full Chern phase diagram is shown in Fig.~\ref{fig:ncl_edge_states}(h). Owing to the large scale of $J/t$, the diagram extends up to $\lambda_{\mathrm{KM}}/t = 1$. Consistent with the band structures, the phase diagram exhibits three topological regions with $\mathcal{C}=\pm 1$ at small KMSOC. As $\lambda_{\mathrm{KM}}/t$ increases, two of these regions are suppressed, while a distinct topological phase $\mathcal{C}=2$ emerges and coexists with the topmost $\mathcal{C}=-1$ phase. This behavior contrasts sharply with the subcritical regime, where increasing KMSOC completely eliminates all Chern phases.

\section{Discussion and conclusion}
\label{sec:summary}

In this work, we investigate edge states of the kagome lattice. This includes both pristine and topological edge states. The pristine states vary strongly with boundary details and are thus labeled trivial. In contrast, topological states remain robust across different terminations and are connected to a bulk invariant. Another important difference is that trivial edge states can be continuously removed without closing the bulk gap, whereas topological edge states cannot. Using a tight-binding framework, we carry out a systematic study of how factors like lattice termination, Kane--Mele and Rashba SOC, Zeeman field, and non-coplanar magnetic order affect the appearance and localization of edge modes. In the pristine limit, we establish that all edge states are trivial and their existence is highly sensitive to termination details:  although most terminations guarantee physical edge modes, the flat termination results in a complete absence of localized edge modes. Based on termination, there can also be valley mixing, leading to a Dirac point-like dispersion at $\Gamma$.

Introducing Kane--Mele SOC, we find that a significantly modified band structure with opened bulk gaps at the K and $\Gamma$ points. When the Fermi level lies within these gaps, the system enters a quantum spin Hall insulating phase. This leads to a time-reversal-protected bulk topology, which guarantees spin-polarized, counter-propagating helical edge states in the slab spectrum, irrespective of termination. Unlike pristine edge states, these helical modes are immune to backscattering due to time-reversal symmetry. A $\mathbb{Z}_2$ index quantifies the bulk topology, as confirmed through the odd winding of Wilson-loop eigenvalue phases across half of the Brillouin zone. We also find that termination-dependent trivial edge states may still also exist in the spectrum, which are not captured by the $\mathbb{Z}_2$ index.

In contrast, we establish that the combination of a Zeeman field, mimicking a ferromagnetic configuration, with Rashba SOC drives the system into a quantum anomalous Hall phase under proper Fermi level tuning. This leads to two Chern insulating phases, $\mathcal{C}=2$ and $\mathcal{C}=1$. The bulk Chern number results in an equal number of co-propagating chiral edge states at the boundary. The bulk gap associated with the $\mathcal{C}=1$ phase is relatively narrow, which limits the resolvability of the edge states. Small amounts of Kane--Mele SOC significantly widen this gap, and flip $\mathcal{C}=1$ to $\mathcal{C}=-1$, reversing the direction of chiral states. We further find that tuning of Kane--Mele coupling destroys the $\mathcal{C}=-1$ phase, while $\mathcal{C}=2$ phase moves to a different filling.

In addition to a Zeeman field, a non-coplanar magnetic texture also leads to Chern phases since it yields a nonzero scalar spin chirality. Here, the canting angle of magnetic moments defines a critical exchange field at which the band Chern number changes and with that the edge state spectrum, creating two distinct exchange-field regimes. Regardless of this scale, we find that the Kane--Mele SOC modifies the bulk bands more strongly in the presence of a magnetic texture than Rashba SOC. At subcritical exchange fields, small Kane--Mele SOC yields two phases with $\mathcal{C}=-2$ and $\mathcal{C}=-1$.
Although $\mathcal{C}=-2$ phase can be obtained with a magnetic texture alone, the $\mathcal{C}=-1$ phase requires Kane--Mele SOC. This landscape changes at supercritical fields, where three topological gaps 
open up with $\mathcal{C}=\pm 1$ for small amounts of Kane--Mele SOC. As SOC is increased, the Chern phases around $\mu/t=\pm 2$ disappear, and a new phase $\mathcal{C}=2$ appears in the vicinity of $\mu/t= 0$. This is unlike the subcritical regime, where increasing SOC completely kills the Chern insulating phase.  

To connect our numerical findings with real materials, we select model parameters that both clearly showcase the edge states and remain consistent with experimental values. In real materials, the SOC-driven Dirac gap is typically an order of magnitude smaller than the bandwidth~\cite{ye.kang.18,yin.ma.20}, which is consistent with our chosen SOC parameters. RSOC can be further tuned using engineered heterostructures~\cite{chen.chen.23}. In fact, similar parameter values as ours have been used in previous tight-binding studies~\cite{zhao.wang.21,sun.chen.22,liu.zhang.09}. When discussing chiral edge states, we consider an effective Zeeman parameter larger than the Kane--Mele SOC scale, since a high Kane--Mele SOC enhances the helical character. This is realistic for strong kagome magnets~\cite{ye.kang.18,yin.ma.20}, such as Fe$_3$Sn$_2$, where the exchange splitting is large~\cite{ye.kang.18}. Additionally, previous studies of kagome Kondo lattice models have explored a broad range of local exchange coupling strengths~\cite{taillefumier.canals.06,barros.venderbos.14,tran.nguyen.21}, including the representative values used in this work. The edge states may also be affected by finite-size effects, leading to inter-edge hybridization. To avoid this, we fix a sufficiently large slab length of $L/a = 20$. To put this into perspective, this value is much larger than the ratio of the edge-state decay length to the lattice constant in the kagome magnet FeGe~\cite{yin.jiang.22}.

In summary, we establish kagome lattice as a versatile platform for engineering tunable edge states. We identify four distinct lattice terminations, which can host up to five edge states. The edge state spectra depend strongly on the termination, possibly explaining distinct edge and surface features observed for different cleaves or terraces~\cite{yin.jiang.22,mazzola.enzer.23}. Certain terminations suppress edge modes or induce valley mixing, with implications for surface and interface engineering. The pristine system hosts trivial edge states embedded in a gapless bulk spectrum, which can be gapped by SOC. In particular, Kane--Mele SOC generates a pair of helical edge modes for in-gap fillings, independent of termination. Such helical modes are expected in nonmagnetic kagome metals such as CsV$_3$Sb$_5$~\cite{ortiz.teicher.20,hu.teicher.22} and KV$_3$Sb$_5$~\cite{ortiz.sarte.21}. When time-reversal symmetry is broken, either by ferromagnetism or non-coplanar magnetic order, bulk Chern gaps emerge with gapless chiral edge modes. Ferromagnetism requires Rashba SOC to open topological gaps, a combination naturally realized in magnetic kagome materials such as Co$_3$Sn$_2$S$_2$~\cite{howard.jiao.21}, TbMn$_6$Sn$_6$~\cite{yin.ma.20}, and FeGe~\cite{yin.jiang.22}. By contrast, non-coplanar magnetic order, as in Mn$_3$Sn~\cite{li.koo.23,low.ghosh.25}, MnBi$_2$Te$_4$~\cite{yang.huang.25}, and Co$M_3$S$_6$ ($M$ = Nb, Ta)~\cite{takagi.takagi.23}, generates distinct Chern phases depending on the exchange field. In both scenarios, Kane--Mele SOC provides an additional control knob, enabling tuning or destruction of Chern phases. Our results are directly relevant for surface engineering, tailored edge transport, or device-oriented topological functionality in kagome-based materials.

\begin{acknowledgments}
S.S. acknowledges the financial support provided by the Polish National Agency for Academic Exchange NAWA  under the Programme STER-Internationalisation of doctoral schools, Project no BPI/STE/2023/1/00027/U/00001. 
S.S. is also grateful for the hospitality and support provided by the Department of Physics and Astronomy at Uppsala University during the early stages of this work.  
S.S. and A.P. further acknowledge support from the National Science Centre (NCN, Poland) under Project No.~2021/43/B/ST3/02166.
A.B.S. acknowledges financial support from the Swedish Research Council (Vetenskapsr\aa det) Grant No.~2022-03963.
We gratefully acknowledge Polish high-performance computing infrastructure PLGrid (HPC Center: ACK Cyfronet AGH) for providing computer facilities and support within computational grant no. PLG/2024/017875 and PLG/2025/018954.
\end{acknowledgments}

\bibliography{biblio.bib}


\clearpage
\newpage

\onecolumngrid

\begin{center}
  \textbf{\Large Supplemental Material}\\[.3cm]
  \textbf{\large Kagome edge states under lattice termination,\\[.1cm] spin--orbit coupling, and magnetic order}\\[.3cm]
  Sajid Sekh$^{1}$, Annica M. Black-Schaffer$^{2}$, and Andrzej Ptok$^{1}$\\[.2cm]
  {\itshape
	$^{1}$Institute of Nuclear Physics, Polish Academy of Sciences, W. E. Radzikowskiego 152, PL-31342 Kraków, Poland\\[.1cm]
    $^{2}$Department of Physics and Astronomy, Uppsala University, Box 516, S-751 20 Uppsala, Sweden
  }
\end{center}

\setcounter{equation}{0}
\renewcommand{\theequation}{SE\arabic{equation}}
\setcounter{figure}{0}
\renewcommand{\thefigure}{SF\arabic{figure}}
\setcounter{section}{0}
\renewcommand{\thesection}{SM\arabic{section}}
\setcounter{table}{0}
\renewcommand{\thetable}{ST\arabic{table}}
\setcounter{page}{1}

\vspace{2cm}

\begin{figure*}[h]
	\includegraphics[width=0.5\linewidth]{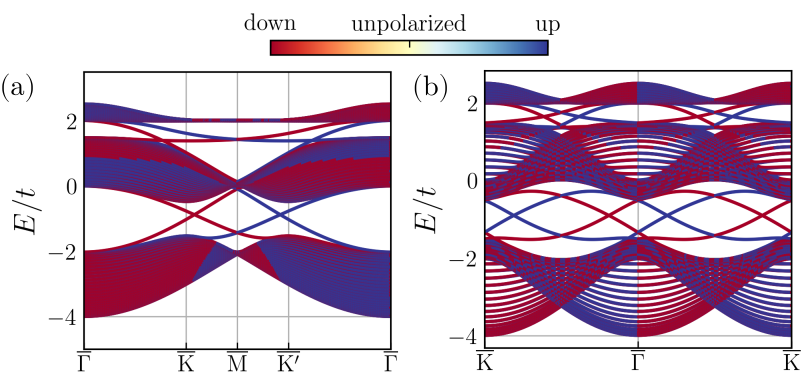}
	\caption{Slab kagome bands in the QSH phase corresponding to Fig.~\ref{fig:soc-bands} in the main text, shown for two different terminations: (a) armchair-flat and (b) zigzag-cove. The color scale represents the spin polarization of the bands. Parameters are $\mu/t = 0$, $\lambda_{\mathrm{KM}}/t = 0.15$, and $t = 1$.}
	\label{fig.sm:sp-ldos}
\end{figure*}

\begin{figure*}[hb]
	\includegraphics[width=\linewidth]{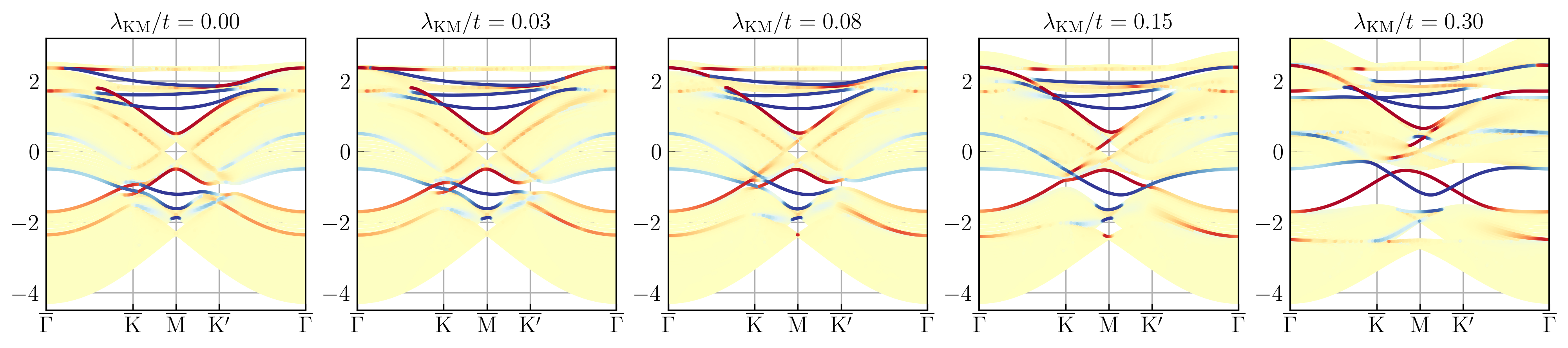}
	\caption{Evolution of slab kagome bands with non-coplanar magnetic order [see Fig.~\ref{fig:ncl_edge_states} in the main text] with increasing Kane--Mele SOC value $\lambda_{\textrm{KM}}/t$. Color indicates expectation of the position operator. Parameters are $\mu/t=0, J/t=0.5$, and $t=1$.}
	\label{fig.sm:qah2qsh}
\end{figure*}

In this Supplemental Material, we present additional results:
\begin{itemize}
\item Fig.~\ref{fig.sm:sp-ldos} -- Slab kagome bands in the QSH phase.
\item Fig.~\ref{fig.sm:qah2qsh} -- Role of KMSOC on the slab kagome bands.
\item Sec.~\ref{sec.sm2} -- Discussion of the flat termination.
\end{itemize}

\newpage

\section{Comment on flat termination}
\label{sec.sm2}

\begin{figure}[!b]
        \includegraphics[width=0.7\linewidth]{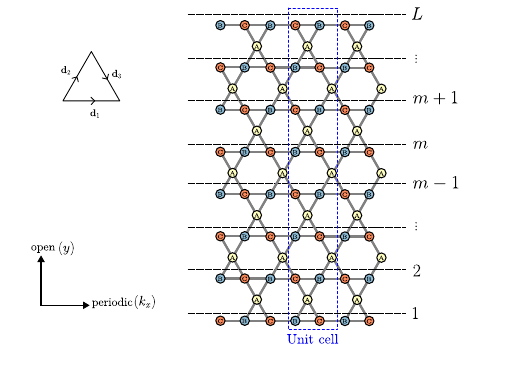}
	\caption{Kagome slab with with a flat edge termination at $m=1,L$. The finite direction of the slab is labeled by $y=1,2,\cdots,m,\cdots,L$, while $k_x$ denotes the periodic direction. Next to the slab, the nearest-neighbor vectors $\mathbf{d}_{1,2,3}$ are shown.}
	\label{fig.sm:flat_kagome}
\end{figure}

In this SM we provide a proof that the flat edge termination of the kagome lattice cannot contain any edge state in the pristine case.
We start by setting up the discrete tight-binding form of the Schr\"odinger equation~\cite{neto.guinea.81}
\begin{align} \label{eq:sm.sch_eq}
E\psi_{\alpha} = \epsilon_{\alpha} \psi_{\alpha} + \sum_{i\neq \alpha} t_{\alpha i} \, \psi_i,
\end{align}
where $\alpha=(A,B,C)$ denotes sites, $i$ refers to NN sites of $\alpha$, and $t_{\alpha i}$ describes NN hopping integral between $\alpha$ and $i$-th site. Upon setting $\epsilon_{\alpha}=0$ for convenience, we can write Eq.~\eqref{eq:sm.sch_eq} for each sublattice site as
\begin{align}
(E/t) \, \psi_{A,m} &= \psi_{B,m} + \psi_{C,m} + e^{iq_2} \psi_{B,m+1} + e^{-iq_3}\psi_{C,m+1}, \\ \nonumber
(E/t) \, \psi_{B,m} &= \psi_{A,m} + \psi_{C,m} + e^{-iq_2} \psi_{A,m-1} + e^{-iq_1}\psi_{C,m}, \\ \nonumber
(E/t) \, \psi_{C,m} &= \psi_{A,m} + \psi_{B,m} + e^{iq_3} \psi_{A,m-1} + e^{iq_1}\psi_{B,m}.
\end{align}
Here, we have defined $q_i = k_x \, d_{ix}, \, (i=1,2,3)$ with $d_{ix}$ being $x$ component of the NN vector $\mathbf{d}_i$, see Fig.~\ref{fig.sm:flat_kagome}. This gives $q_1=k_x/2$, $q_2=k_x/4$, and $q_3=-k_x/4$.

For the flat edge termination, we now write the boundary condition for the $m=1$ edge, see Fig.~\ref{fig.sm:flat_kagome}, as
\begin{align} \label{eq:sm.bc_bottom}
(E/t) \, \psi_{A,1} &= \psi_{B,1} + \psi_{C,1} + e^{iq_2} \psi_{B,2} + e^{-iq_3}\psi_{C,2}, \\ \nonumber
(E/t) \, \psi_{B,1} &= \psi_{A,1} + \psi_{C,1} + e^{-iq_1}\psi_{C,1}, \\  \nonumber
(E/t) \, \psi_{C,1} &= \psi_{A,1} + \psi_{B,1} + e^{iq_1}\psi_{B,1}.
\end{align}
and for the $m=L$ edge
\begin{align} \label{eq:sm.bc_top}
(E/t) \, \psi_{B,L} &= \psi_{C,L} + e^{-iq_1}\psi_{C,L} + e^{-iq_2} \psi_{A,L-1}, \\  \nonumber
(E/t) \, \psi_{C,L} &= \psi_{B,L} + e^{iq_1} \psi_{B,L} + e^{iq_3}\psi_{A,L-1}.
\end{align}
To solve the above equations, we take the evanescent mode Ansatz 
\begin{align} \label{eq:sm.ansatz}
\psi_{\alpha} = \lambda^m \phi_{\alpha}.
\end{align}
With this expression, $|\lambda|>1$ or $|\lambda|<1$ indicate exponentially localized mode pinned at the top or bottom edge, respectively. If $|\lambda|=1$ however, the wave function does not depend on $m$, which signifies an extended bulk-like character. Substituting Eq.~\eqref{eq:sm.ansatz} in Eq.~\eqref{eq:sm.bc_bottom} and \eqref{eq:sm.bc_top} yields
\begin{align}
(E/t) \, \phi_{A} &= (1 + \lambda e^{iq_2}) \, \phi_{B} + (1 + \lambda e^{-iq_3}) \, \phi_{C}, \\  
\label{eq:sm.bc_bottom_B}
(E/t) \, \phi_{B} &= \phi_{A} + (1 + e^{-iq_1}) \, \phi_{C}, \\ 
\label{eq:sm.bc_bottom_C}
(E/t) \, \phi_{C} &= \phi_{A} + (1 + e^{iq_1}) \, \phi_{B}.
\end{align}
and
\begin{align}
\label{eq:sm.bc_top_B}
(E/t) \, \phi_{B} &= \lambda^{-1} e^{-iq_2} \phi_{A} + (1 + e^{-iq_1}) \, \phi_{C}, \\ 
\label{eq:sm.bc_top_C}
(E/t) \, \phi_{C} &= \lambda^{-1} e^{iq_3} \phi_{A} + (1 + e^{iq_1}) \, \phi_{B}. 
\end{align}
Next, adding Eq.~\eqref{eq:sm.bc_bottom_B} and \eqref{eq:sm.bc_bottom_C} and Eq.~\eqref{eq:sm.bc_top_B} and \eqref{eq:sm.bc_top_C} gives us
\begin{align} 
\label{eq:sm.final1}
(E/t) \, (\phi_{B}+\phi_{C}) &= 2\phi_{A} + (1 + e^{iq_1}) \, \phi_{B} + (1 + e^{-iq_1}) \, \phi_{C} \\ 
\label{eq:sm.final2}
(E/t) \, (\phi_{B}+\phi_{C}) &= \lambda^{-1} (e^{-iq_2} + e^{iq_3}) \, \phi_{A} + (1 + e^{iq_1}) \, \phi_{B} + (1 + e^{-iq_1}) \, \phi_{C}
\end{align}
Then, comparing Eq.~\eqref{eq:sm.final1} and ~\eqref{eq:sm.final2} we obtain
\begin{align}
\lambda = \frac{e^{-iq_2} + e^{iq_3}}{2}
\end{align}
Finally, using the values of $q_2$ and $q_3$ we arrive at $\lambda=e^{-ik_x/4}$ or $|\lambda|=1$. This indicates no localized mode solution exists for flat termination, which is in line with the results in Sec.~\ref{sec:slab_termination} in the main text.

\end{document}